\documentclass[preprint,12pt]{elsarticle}

\usepackage{amssymb}
\usepackage{amsmath}

\journal{Physica D: Nonlinear Phenomena}

\usepackage{setspace}
\usepackage[T1]{fontenc}
\usepackage[utf8]{inputenc}
\usepackage{xcolor}
\usepackage{array}
\usepackage{mathtools}
\usepackage{mathrsfs}
\usepackage{mdframed}
\usepackage{multirow} 
\usepackage{multicol} 
\usepackage{subcaption}
\usepackage{comment}
\usepackage{dirtytalk}
\usepackage{empheq}
\usepackage[colorlinks,allcolors=black]{hyperref}

\begin{document}

\begin{frontmatter}
	


\title{Modeling late-time sensitivity to initial conditions in Boussinesq Rayleigh-Taylor turbulence}


\author[affilENS,affilCEA]{Sébastien Thévenin} 
\author[affilCEA,affilLMCE]{Benoît-Joseph Gréa} 


\affiliation[affilENS]{organization={Université Paris Saclay, Université Paris Cité, ENS Paris Saclay, CNRS, SSA, INSERM, Centre Borelli},
	postcode={F-91190},
	city={Gif-sur-Yvette},
	country={France}}

\affiliation[affilCEA]{organization={CEA, DAM, DIF},
	postcode={F-91297},
	city={Arpajon},
	country={France}}

\affiliation[affilLMCE]{organization={Université Paris-Saclay, CEA, LMCE},
	postcode={F-91680},
	city={Bruyères-le-Châtel},
	country={France}}

\begin{abstract}
	
	This article sheds light on the late-time influence of initial conditions in Boussinesq Rayleigh-Taylor turbulence using an approach combining direct numerical simulations, machine learning and theory. The initial conditions are characterized by four non-dimensional numbers describing the statistical properties of random-phase multi-mode perturbations of an initial diffuse interface. Based on high-fidelity data, a surrogate physics-informed neural network is used to extrapolate the dynamics to very late times and unseen initial conditions, beyond the reach of simulations. This enables uncertainty and global sensitivity analyses to be carried out, revealing the influence of initial conditions in the late-time regime. While the results support the idea of a universal self-similar growth rate, the virtual time origin is found to be strongly sensitive to the initial Reynolds, perturbation steepness and bandwidth numbers. An analytical model based on the phenomenology of Rayleigh-Taylor mixing layers explains most of this dependency, and provide accurate predictions for the virtual time origin. It turns out that when the initial perturbation reaches nonlinear saturation earlier, the mixing layer also re-accelerates earlier, while the virtual time origin is larger.
	
	
\end{abstract}

\begin{graphicalabstract}
	\includegraphics[width=\linewidth]{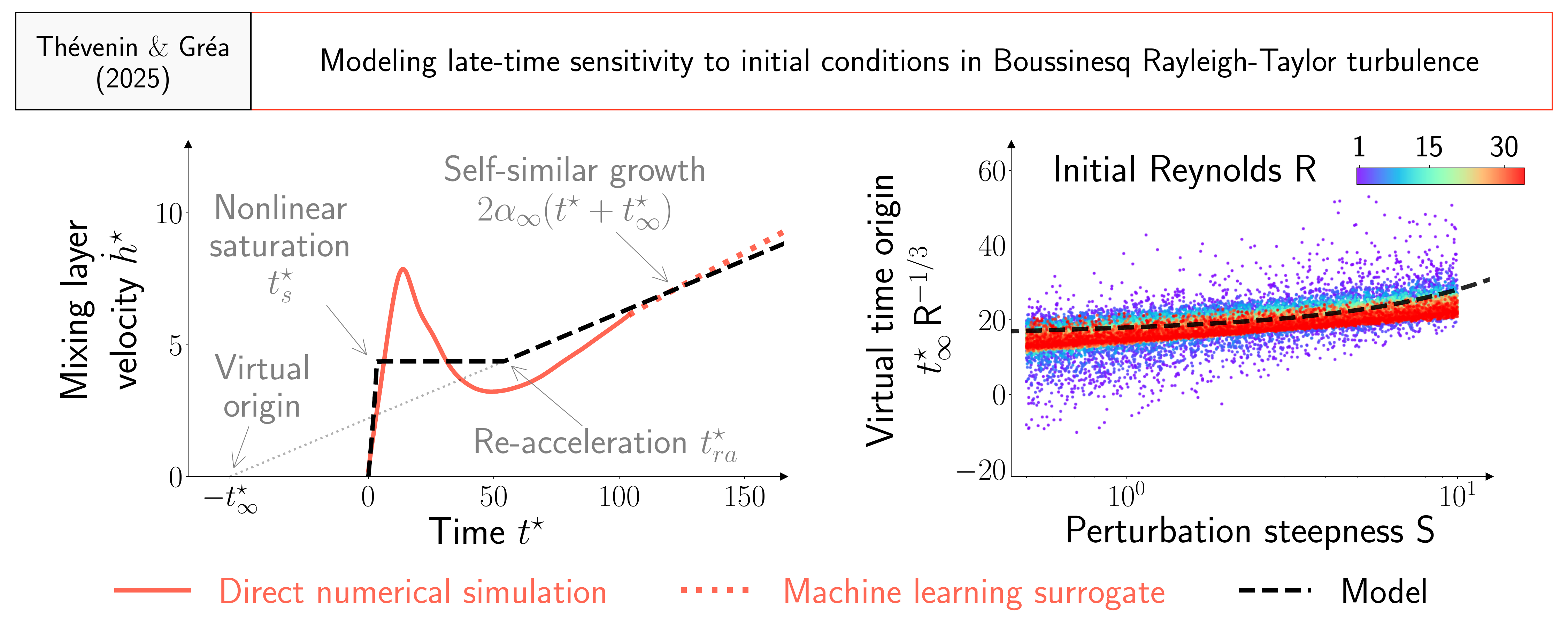}
\end{graphicalabstract}

\begin{highlights}
	\item Late-time sensitivity to initial conditions is imprinted in the virtual origin
	\item The initial Reynolds, perturbation steepness and bandwidth numbers are key factors
	\item An analytical model is built to predict the virtual time origin
	\item Early nonlinear saturation leads to early re-acceleration and larger virtual origin
	\item A work combining simulations, physics-informed machine learning, and theory
\end{highlights}


\end{frontmatter}

\section{Introduction}

The Rayleigh-Taylor (RT) instability is encountered in various fields \citep{Zhou2017_I,Zhou2017_II}, including geophysics \citep{davey_rotating_1981,khomenko_rayleigh-taylor_2014,ravichandran_mammatus_2020}, astrophysics \citep{bell_direct_2004,ryutova_observation_2010,mishra_evolution_2018} and engineering applications such as inertial confinement fusion \citep{lindl_development_1995,betti_inertial-confinement_2016,remington_rayleightaylor_2019,zhou_instabilities_2025}. It appears at the interface separating two fluids of different densities under the effect of an acceleration directed toward the lighter fluid, in the opposite direction of the density gradient \citep{Rayleigh_1882,Taylor_1950,sharp_overview_1984,boffetta_incompressible_2017}. This is the case, for instance, when the heavy fluid is above the light fluid in a gravity field. The destabilizing acceleration amplifies any perturbation of the interface, leading to the development of a mixing layer. 
The shear created at the interface later triggers secondary instabilities, marking the start of the transition to turbulence. Since the pionneering experiments and simulations of Read \citep{read_experimental_1984} and Youngs \citep{youngs_numerical_1984}, it has been well established that the height $h$ of the mixing layer upper half evolves toward a quadratic solution at late times,
\begin{equation}\label{eq:self_similar_solution}
	h(t)=\alpha_\infty \mathcal{A}g(t+t_\infty)^2,
\end{equation}
which is proportional to the Atwood number $\mathcal{A}=(\rho_h-\rho_l)/(\rho_h+\rho_l)$, characterizing the density contrast between the heavy and light fluids of densities $\rho_h$ and $\rho_l$, and the acceleration $g$, with $t$ the time that is equal to zero at the moment of the disturbance. This evolution is supported by theoretical arguments involving self-similarity \citep{ristorcelli_rayleightaylor_2004} or mass flux and energy budgets \citep{cook_mixing_2004}. It is also solution to the buoyancy-drag equation, a simple model commonly used to describe RT flows throughout all their stages, which can be derived from first principles \citep{ramshaw_simple_1998,dimonte_spanwise_2000,grea_rapid_2013,schilling_buoyancysheardrag-based_2020}.

An important fundamental question in the study of RT flows is whether the self-similar parameters $\alpha_\infty$ and $t_\infty$ depend on initial conditions. If so, this has far-reaching implications for the development of turbulence models. On the one hand, their calibration relies in part on the reproduction of the late-time solution, Eq. \eqref{eq:self_similar_solution}, which makes the model coefficients dependent on the self-similar parameters. On the other hand, the models themselves should integrate this dependency to be able to describe the transition and bridge the gap between the early phases of RT instability and fully developed turbulence. Many studies are already moving in this direction, by modifying the model's structure and closure \citep{schilling_turbulent_2017,xie_intermittency_2025}, its initialization \citep{rollin_generating_2013} or the choice of variables and solutions to be reproduced \citep{thevenin_al_JFM_2025}.

The self-similar growth rate $\alpha_\infty$ has long been a central concern, as large differences were reported in the literature \citep{Dimonte_alpha_group_2004}. Values measured in experiments, up to $\sim0.070$, were indeed much higher than those obtained in numerical simulations, around $\sim0.020-0.025$. Today, there seem to be a number of reasons for these differences. First, it is very difficult to perform RT experiments with controlled initial conditions and without adding extra effects such as tilt, shear or variable accelerations \citep{banerjee_rayleigh-taylor_2020}, complicating a proper comparison with simulations. Recent experiments that achieve this compare very well with values found in simulations \citep{roberts_effects_2016}. Second, the bounded nature of experiments and numerical simulations can induce large-scale confinement of the flow structures, which alters the dynamics. This may occur earlier if scales close to the lateral size of the system are present in the initial disturbance spectrum, which may be associated to the mode competition case of Dimonte \citep{Dimonte_dependence_2004,ramaprabhu_numerical_2005}. As pointed out by Cabot and Cook \citep{cabot_reynolds_2006}, another reason stems from the evaluation of $\alpha_\infty$, which was often calculated as $h(t)/(\mathcal{A}gt^2)$. This estimate, which neglects the linear term of the self-similar solution, converges very slowly to its asymptotic value, resulting in overestimates if evaluations are made too early. Finally, the definition of the height $h$ can also lead to different values for $\alpha_\infty$. Many works now support the idea of a universal value for the asymptotic growth rate in the case of unbounded RT, from large direct numerical simulations \citep{cabot_reynolds_2006,morgan_parametric_2020,briard_growth_2022,thevenin_al_JFM_2025} and well-controlled experiments \citep{roberts_effects_2016} to theoretical arguments \citep{poujade_growth_2010,soulard_large-scale_2015}.

The virtual origin, expressed as the time $t_\infty$ in Eq. \eqref{eq:self_similar_solution}, has a somewhat more obscure meaning. While it is a parameter characterizing the late asymptotic regime, it seems instinctively linked to what happens during the early stages of the instability, and hence to the initial conditions. However, unlike the growth rate, it is still poorly understood and has not received much attention, despite having a significant impact on modeling and predicting the RT dynamics. Perhaps this is because it is very difficult to estimate, both in experiments and in simulations. Indeed, it seems that the only way to compute it is to use Eq. \eqref{eq:self_similar_solution} and its time derivative $\dot{h}$, which gives an expression that converges very slowly to its asymptotic value. 
As a result, experiments and simulations must run for a sufficiently long time, without confining the large scales. In practice, this is very difficult to achieve, as it would require initializing the perturbation spectrum with very small wavelengths, which implies very sophisticated experiments or very high-resolution simulations.

This challenge is addressed in this work with the help of a machine learning surrogate model capable of realistic time extrapolation, enabling the calculation of converged estimates of the self-similar parameters. This allows to investigate the late-time sensitivity of RT to the initial conditions, focusing especially on the virtual origin. The study considers the case of low density contrast and miscible fluids under the Boussinesq approximation, such that the mixing layer maintains a top/bottom statistical symmetry \citep{Ramaprabhu_experimental_2004,livescu_numerical_2013} and its size is twice the height $h$. The geometry is three-dimensional and considered unbounded, the interface is planar and the acceleration is constant. The initial conditions are here parameterized by four non-dimensional numbers, as in Thévenin \textit{et al.} \citep{thevenin_al_JFM_2025}. These represent the statistical properties of a diffuse interface with random-phase multi-mode perturbations taking the form of top-hat and annular Fourier spectra.

The paper is organized as follows. The physical configuration is first presented in details, and the numerical and data-driven tools needed to carry out the study are introduced. In the part that follows, we recall the typical behaviors of miscible RT-unstable mixing layers, classifying them into three groups. A variance-based global sensitivity analysis is then performed in three different domains, revealing the initial conditions parameters that most influence the self-similar regime. Finally, we propose a simple phenomenological model connecting the early and late stages of the RT instability. This provides an analytical relationship between the virtual time origin and the nonlinear saturation time of the initial perturbation.

\section{Physical, numerical and data-driven setups}

\subsection{Rayleigh-Taylor flow configuration}\label{subsec:flow_configurations_and_IC}

At low density contrast, the flow induced by the RT instability can be represented by the velocity $\boldsymbol{u}(\boldsymbol{x},t)=(u_x,u_y,u_z)^T$ and the concentration of heavy fluid $c(\boldsymbol{x},t)=(\rho-\rho_l)/(\rho_h-\rho_l)$, in a Cartesian frame of coordinates $\boldsymbol{x}=(x,y,z)^T$ and at a time $t$. Under the Boussinesq approximation, density differences only need to be accounted for in the buoyancy term. The flow is therefore described by the incompressible Navier-Stokes equations, supplemented by a transport equation for the scalar concentration field,
\begin{subequations}\label{eqs:Navier-Stokes-Boussinesq}
	\begin{align}
		&\boldsymbol{\nabla}\cdot\boldsymbol{u}=0, \label{eq:NS_continuity}\\
		&\partial_t \boldsymbol{u} +\boldsymbol{u}\cdot\boldsymbol{\nabla}\boldsymbol{u} = -\boldsymbol{\nabla} \Pi -2\mathcal{A}g c \,\boldsymbol{e}_z + \nu\nabla^2\boldsymbol{u}, \label{eq:NS_momentum}\\
		&\partial_t c +\boldsymbol{u}\cdot\boldsymbol{\nabla}c = \kappa \nabla^2 c. \label{eq:NS_transport}
	\end{align}
\end{subequations}
In these equations, $\Pi=p/\rho_0$ is the reduced pressure, with $\rho_0=(\rho_h+\rho_l)/2$ the constant mean density and $p(\boldsymbol{x},t)$ a deviation from the hydrostatic pressure, $\boldsymbol{e}_z$ is the unit basis vector in the vertical direction, $\nu$ is the fluids' kinematic viscosity and $\kappa$ is the molecular diffusion coefficient. In this study, we consider a unit Schmidt number, $Sc=\nu/\kappa=1$.

The fluids are assumed to be initially at rest, with zero velocity, and the interface is diffuse and disturbed, so that the initial conditions write
\begin{subequations}\label{eqs:init_u_C_fields}
	\begin{align}
		&\boldsymbol{u}(\boldsymbol{x},0)=\boldsymbol{0}, \label{subeq:init_u}\\
		&c(\boldsymbol{x},0)=\dfrac{1}{2}+\dfrac{1}{2}\tanh\left(\dfrac{3\,[z-\eta(x,y)]}{\delta}\right). \label{subeq:init_C}
	\end{align}
\end{subequations}
In the above definition, $\delta$ is the initial diffusive thickness of the interface, which is equal to the initial mixing zone size, defined in Eq. \eqref{def:mixing_zone_size}, in the absence of disturbance, hence the factor $3$ in the hyperbolic tangent numerator. The zero-mean interface perturbation $\eta(x,y)$ is chosen to be a superposition of $2\pi$-periodic Fourier modes, defined such that the resulting two-dimensional spectrum has a top-hat, annular shape, and the power spectral density has a rectangular shape with a constant amplitude $a_0$ for the modes of the spectral band. In particular, $\eta(x,y)$ has a variance $\eta_0^2$, with $\eta_0$ the root mean square (rms) amplitude of the perturbation, and only the modes of wavenumber $k\in[k_0-\frac{1}{2}\Delta k,k_0+\frac{1}{2}\Delta k]$ have a non-zero amplitude and a random phase. The wavenumber is here defined as $k=\sqrt{k_x^2+k_y^2}$, based on the components of the two-dimensional wavevector $\boldsymbol{k}=(k_x,k_y)^T$. The spectral band is therefore defined by a mean wavenumber $k_0$ and a bandwidth $\Delta k$, which is related to the variance through $\eta_0^2=a_0 \Delta k$ due to the rectangular shape of the Fourier power spectrum.

The initial conditions are illustrated and described in greater details in the associated paper and database \citep{thevenin_al_JFM_2025,thevenin_database_2025}, and can be characterized by four non-dimensional numbers,
\begin{equation}\label{def:RBSD}
	\mathsf{R}=\dfrac{\sqrt{\mathcal{A}gk_0}}{\nu k_0^2} \quad;\quad \mathsf{B}=\dfrac{\Delta k}{k_0} \quad;\quad \mathsf{S}=\eta_0 k_0 \quad;\quad \mathsf{D}=\delta k_0.
\end{equation}
These are respectively an initial Reynolds number, defined as the ratio between the classical and viscous asymptotic linear growth rates of the RT instability \citep{Duff_effects_1962}; a bandwidth number, quantifying the number of modes having a non-zero initial amplitude and characterizing the importance of interactions between modes; a perturbation steepness number, describing how flat the initial interface is and characterizing the onset of nonlinearity; and a diffusive thickness number, describing how diffuse the initial interface is and reducing the linear growth rate in favor of an initially diffusive growth.

Although these numbers do not fully describe the initial conditions, as the random phases are not known, they are sufficient to describe the statistical properties of the initial interface.

\subsection{Definitions of the late-time dynamical properties} \label{subsec:late_time_dynamics}

As a reminder, RT mixing layers maintain a top/bottom statistical symmetry when the density contrast is low \citep{Ramaprabhu_experimental_2004,livescu_numerical_2013}. The mixing layer half height can be defined as
\begin{equation}\label{def:mixing_zone_size}
	h(t)=3\int_{-\infty}^{+\infty} \overline{C}(z,t) \left(1-\overline{C}(z,t)\right)\,\mathrm{d}z,
\end{equation}
following the convention introduced by Andrews and Spalding \citep{andrews_simple_1990}, which assumes that the horizontally-averaged, vertical concentration profile $\overline{C}(z,t)=\iint c(\boldsymbol{x},t) \,\mathrm{d}x\,\mathrm{d}y$ is piecewise linear. Although this is not true in general, it provides a definition robust to noise and small fluctuations.

The expansion velocity is defined as the first time derivative of the height, which is equal to
\begin{equation}\label{def:expansion_velocity}
	\dot{h}(t)=3\int_{-\infty}^{+\infty} \left(\kappa \,\partial_{zz}\overline{C}(z,t)-\partial_z \overline{u_z^\prime c^\prime}(z,t)\right)\left(1-2\,\overline{C}(z,t)\right)\,\mathrm{d}z,
\end{equation}
after injecting the horizontally-averaged scalar transport equation \eqref{eq:NS_transport}, and with the quantities $u_z^\prime(\boldsymbol{x},t)=u_z(\boldsymbol{x},t)-\overline{U_z}(z,t)$ and $c^\prime(\boldsymbol{x},t)=c(\boldsymbol{x},t)-\overline{C}(z,t)$ corresponding to the fluctuations of vertical velocity and concentration. The expansion is thus driven by molecular diffusion and convective mass transfer, represented respectively by the first and second terms in the left parenthesis.

From these definitions, it is possible to characterize the late-time growth of the mixing layer with
\begin{equation}\label{def:transient_growth_rate_virtual_origin}
	\alpha(t)=\dfrac{\dot{h}^2(t)}{4\mathcal{A}gh(t)}\underset{t\rightarrow +\infty}{\longrightarrow} \alpha_\infty \quad\quad\text{and}\quad\quad \dfrac{2h(t)}{\dot{h}(t)}-t \underset{t\rightarrow +\infty}{\longrightarrow} t_\infty,
\end{equation}
This estimate for the asymptotic self-similar growth rate converges much faster to $\alpha_\infty$ than the one discussed in the introduction, $h(t)/(\mathcal{A}gt^2)$. On the other hand, the virtual time origin estimate converges very slowly to $t_\infty$. For this reason, it is important to ensure that large-scale confinement does not influence the late-time dynamics.

In the rest of the paper, the quantities previously defined are often renormalized using the buoyant acceleration $\mathcal{A}g$ and the kinematic viscosity $\nu$. The resulting non-dimensional quantities are marked by a $\star$ symbol in the exponent. As this study considers an unbounded RT configuration, this renormalization helps to focus on the behavior of each example and the role of its initial conditions, independently of the dimensional values used. In particular, two examples initialized differently but sharing the same values for the numbers $\mathsf{R}$, $\mathsf{B}$, $\mathsf{S}$ and $\mathsf{D}$ will display similar, statistically overlapping non-dimensional dynamics in the absence of large-scale confinement (see for instance Thévenin \citep{thevenin_contrib_2024}).

\subsection{Surrogate model for direct numerical simulations}\label{subsec:DNS_and_surrogate}

As mentioned in the introduction, it is difficult to evaluate the late-time properties of the mixing layer. This is particularly true for the virtual origin, even with simulations. To tackle this challenge, we build on and expand an existing work \citep{thevenin_al_JFM_2025,thevenin_contrib_2024}, which makes use of a surrogate model capable of performing realistic time extrapolations.

The surrogate model is built with a physics-informed neural network \citep{raissi_physics-informed_2019,karniadakis_physics_informed_2021} trained on direct numerical simulations (DNS) data to reproduce a set of volume-averaged quantities. These include the mixing layer height and expansion velocity, which are parametrized by a set of initial conditions $\mathsf{I}=(\mathsf{R,B,S,D})^T$ and a time $t^\star$. The training data, freely available online on the \textit{French Fluid Dynamics Database} \citep{thevenin_database_2025}, comprises 493 DNS performed with the code \texttt{Stratospec} \citep{viciconte_sudden_2019,Briard_harmonic_2019,Briard_turbulent_2024}. Information regarding numerical resolution and constraints can be found in the paper documenting the database \citep{thevenin_database_2025}.

In addition to being trained on the DNS data, the surrogate is constrained by its architecture and training loss function to comply as closely as possible with several physical properties. These include the quadratic time evolution of $h(t)$, the conservation of kinetic energy and scalar variance, and Gréa's rapid acceleration model \citep{grea_rapid_2013} which frames (without fixing) the values of the growth rate by relating it to the mixing properties. As no data is required to compute the terms of the physics-informed loss function, these are also evaluated for initial conditions and times that are not available in the database. This is what enables the surrogate to make realistic extrapolations at late times, and therefore to estimate the asymptotic self-similar quantities. Besides, its very low computational cost allows to generate a very large number of samples very quickly for various initial conditions. It is important to point out that the values obtained for $\alpha_\infty$ and $t_\infty$ have not been specified or fixed arbitrarily, but rather stem from the learning process.

\section{Typical behaviors of miscible Rayleigh-Taylor mixing layers}\label{sec:behaviors_unbounded_RT}

In this section, we review the expected behaviors of RT mixing layers in a miscible configuration. The aim is to map these behaviors in the space of non-dimensional numbers characterizing the initial conditions, which enables us to roughly anticipate their influence on the late-time parameters describing the mixing layer growth.

\subsection{Examples}

\begin{figure}[!t]
	\centering
	\begin{subfigure}[b]{\linewidth}
		\centering
		\includegraphics[width=\linewidth]{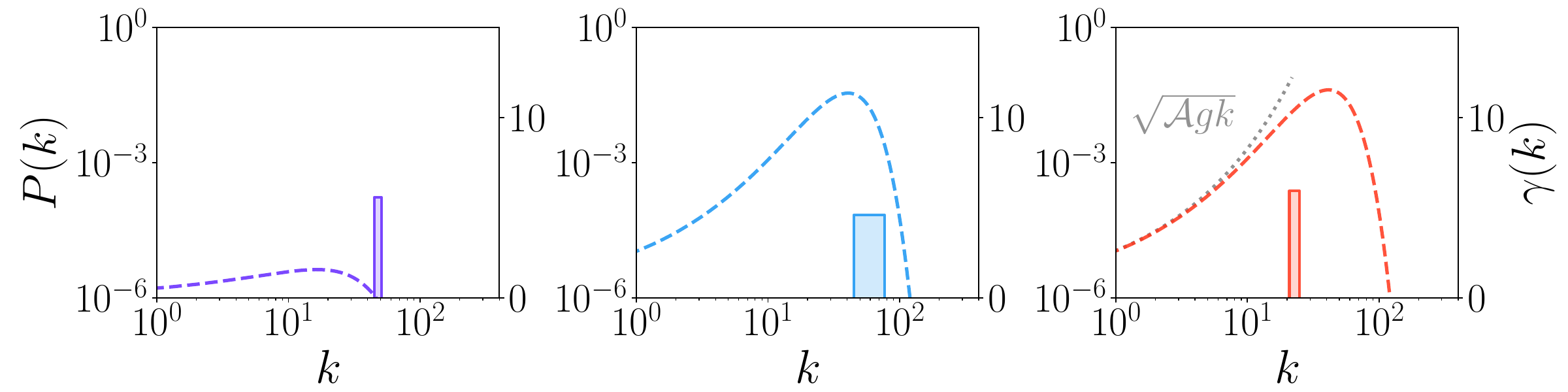}
		\caption{Spectral band of the perturbations (left axis) and linear growth rate profiles (right axis) according to the expression formulated by Duff, Harlow and Hirt \citep{Duff_effects_1962}.}
		\label{subfig:examples_init_dim}
	\end{subfigure}
	\vfill
	\begin{subfigure}[b]{\linewidth}
		\centering
		\includegraphics[width=\linewidth]{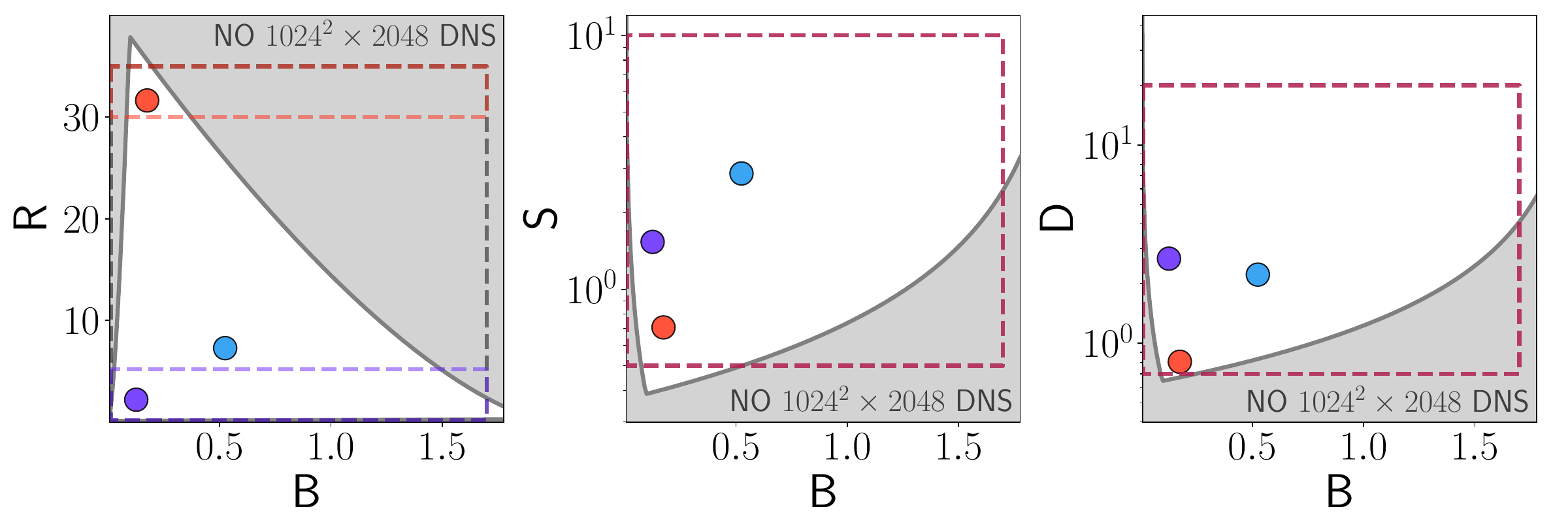}
		\caption{Distribution in the non-dimensional numbers space. The dashed rectangles correspond to the domains in which global sensitivity analyses are performed in Section \ref{sec:late_time_dependence}.}
		\label{subfig:examples_init_nondim}
	\end{subfigure}
	\caption{Initial conditions of the three examples illustrating the typical behaviors of miscible RT mixing layers. Their parameters are given in Table \ref{tab:inits_examples_3_regimes}.}
	\label{fig:examples_init}
\end{figure}

\begin{figure}[t!]
	\centering
	\includegraphics[width=\linewidth]{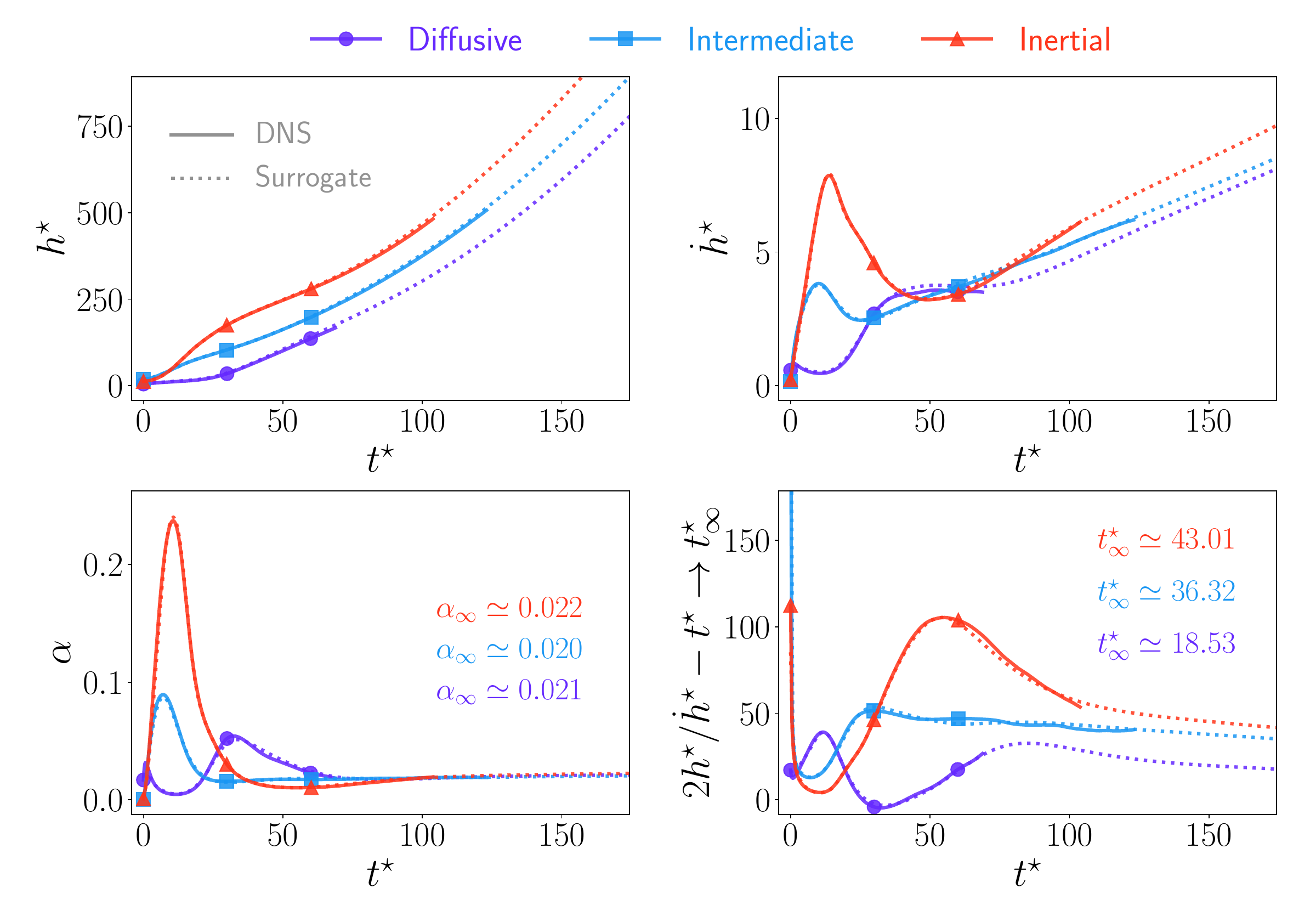}
	\caption{Non-dimensional dynamics of the three examples whose initial conditions are given in Table \ref{tab:inits_examples_3_regimes} and shown in Figure \ref{fig:examples_init}. Solid lines correspond to the DNS and dotted lines to the surrogate model predictions. The surrogate allows to extrapolate the trajectories to later times, and hence to estimate the asymptotic self-similar parameters $\alpha_\infty$ and $t_\infty^\star$. The values shown in the figure correspond to the average over the time window $t^\star\in[160,170]$.}
	\label{fig:examples_dyn}
\end{figure}

Three typical examples are used for illustration, and correspond to cases referred to as \textit{diffusive} (purple), \textit{intermediate} (blue) and \textit{inertial} (red) in the following. Their initial conditions are given in Table \ref{tab:inits_examples_3_regimes} and shown in Figure \ref{fig:examples_init}, while their dynamics is shown in Figure \ref{fig:examples_dyn}. Figure \ref{subfig:examples_init_dim} shows how the perturbation spectral bands are positioned with respect to the linear growth rate profiles according to the model of Duff, Harlow and Hirt \citep{Duff_effects_1962}, whereas Figure \ref{subfig:examples_init_nondim} shows their position in the four non-dimensional space $(\mathsf{R,B,S,D})$.

\bgroup
\def\arraystretch{1.2} 
\setlength{\tabcolsep}{0.55em}
\begin{table}[!b]
	\begin{center}
		\begin{tabular}{|ccccccccc|}
			\hline 
			$\mathsf{R}$ & $\mathsf{B}$ & $\mathsf{S}$ & $\mathsf{D}$ & $k_0$ & $\Delta k$ & $\eta_0$ & $\delta$ & $\mathcal{A}g$ \\
			\hline
			\hline
			$2.196$ & $0.125$ & $1.537$ & $2.668$ & $48$ & $6$ & $0.032$ & $0.056$ & $0.3$ \\
			$7.271$ & $0.525$ & $2.861$ & $2.217$ & $61$ & $32$ & $0.047$ & $0.036$ & $6.75$ \\
			$31.64$ & $0.174$ & $0.708$ & $0.806$ & $23$ & $4$ & $0.031$ & $0.035$ & $6.85$ \\
			\hline
		\end{tabular}
		\caption{Initialization parameters of the 3 DNS used as examples to illustrate the \textit{diffusive}, \textit{intermediate} and \textit{inertial} regimes, respectively from top to bottom. All simulations have an Atwood number $\mathcal{A}=0.05$, a lateral box size equal to $2\pi$, a grid resolution of $1024^2\times2048$ and a viscosity $\nu=7.5\times10^{-4}$.}
		\label{tab:inits_examples_3_regimes}
	\end{center}
\end{table}
\egroup

The purple example first experiences a diffusive growth with an expansion velocity $\dot{h}^\star$ close to zero and remaining fairly monotonic throughout all stages (see Figure \ref{fig:examples_dyn}). It results a small value of the virtual time origin $t^\star_\infty$. On the other hand, the inertial example in red undergoes a strong acceleration right from the start, and is only slowed down around $t^\star\sim15$. At that time, secondary shear instabilities have developed and the mixing layer reaches a local peak in velocity, which can be shown to also coincide with a peak in kinetic energy dissipation. Later, the mixing layer re-accelerates. This non-monotonic evolution leads to a greater virtual time origin. The blue example follows an in-between, intermediate behavior, with a velocity peak that is not very pronounced.

The predictions of the surrogate model look realistic, as shown by the dotted lines in Figure \ref{fig:examples_dyn}. They allow to estimate the asymptotic values for the self-similar growth rate and virtual time origin. The former seems to be roughly constant around $\alpha_\infty\sim0.021$, while the second spans a wider range of values, $t^\star_\infty\in[18,43]$, which indicates a strong dependency on initial conditions.

\subsection{Description based on linear stability}

Behaviors close to the red example occur when the destabilizing buoyant acceleration ($\mathcal{A}g$) is very strong compared to the dissipative mechanisms ($\nu,\kappa$). Such case, referred to as \textit{inertial}, is characterized by a large initial Reynolds number $\mathsf{R}$ and has a linear growth rate that closely resembles the classical inviscid and immiscible one \citep{Rayleigh_1882,Taylor_1950,Chandrasekhar1961}, $\gamma(k)=\sqrt{\mathcal{A}gk}$ (see Figure \ref{subfig:examples_init_dim}). In non-dimensional $\star$ form, the linear growth rate of the mean wavenumber $k_0$ can thus be written as $\gamma_0^\star(\mathsf{R})=\mathsf{R}^{-1/3}$.

In the case of a diffuse interface, this classical linear growth rate is damped, and can rather be expressed as
\begin{equation}\label{eq:linear_growth_rate_nondim}
	\gamma_0^\star(\mathsf{R},\psi)=\mathsf{R}^{-4/3}\left(\sqrt{\dfrac{\mathsf{R}^2}{\psi}+1}-2\right)
\end{equation}
for the mean wavenumber $k_0$ of the perturbation, according to Duff, Harlow and Hirt's model \citep{Duff_effects_1962}. The function $\psi$ represents a damping caused by the initial diffusive thickness $d$ of the interface. At low Atwood number, it can be expressed as $\psi=1+\frac{\sqrt{2}}{\pi}k_0 \,d$ \citep{Duff_effects_1962}, with $d=1/\max(\partial c(\boldsymbol{x},0)/\partial z)$. In the absence of perturbation $(\eta(x,y)=0)$, $d=2\delta/3$ from Eq. \eqref{subeq:init_C}, which introduces a dependency to $\mathsf{D}$. When the perturbation is not zero, the effective diffusive thickness is greater, and can be represented by the initial mixing zone size, $2h(0)$, leading to $d=4h(0)/3$. With various simplifications (see \citep{thevenin_contrib_2024}), we can show from Eq. \eqref{subeq:init_C} that
\begin{equation}\label{eq:initial_mixing_zone_size_h0}
	2 h(0)k_0 \simeq (3/2)\sqrt{\left(\mathsf{D}/3\right)^2+\mathsf{S}^2}\left(1+\tanh^2\left[3\mathsf{S}/\mathsf{D}\right]\right) + \mathsf{D}/2,
\end{equation}
which was shown to be accurate to within $10\%$ compared to the database's DNS \citep{thevenin_contrib_2024}. This introduces a dependency to the steepness number $\mathsf{S}$ that further damps the linear growth rate. Eq. \eqref{eq:initial_mixing_zone_size_h0} leads to $2h(0)=\delta$ when $\mathsf{D}\gg\mathsf{S}$, retrieving the solution in the absence of perturbation, and to $2h(0)=3\eta_0$ when $\mathsf{D}\ll\mathsf{S}$. From Eq. \eqref{eq:linear_growth_rate_nondim}, it is possible to determine the most unstable ($\gamma_0=\gamma_{\text{max}}$) initial conditions as those having an initial Reynolds number equal to
\begin{equation}\label{eq:most_unstable_initial_R}
	\mathsf{R}_{\text{max}}^2 = \psi\left[ 30+8\sqrt{13} \right]
\end{equation}
Initial conditions with Reynolds greater than $\mathsf{R}_{\text{max}}$ are expected to follow inertial behaviors.

In addition, molecular diffusion also has dynamical effects that may become predominant at early times, such as in the purple \textit{diffusive} example. In the absence of disturbance, for instance, the mixing layer is expected to grow as $h(t)=\sqrt{h(0)+6\kappa t}$, by integrating Eq. \eqref{def:expansion_velocity} using Eq. \eqref{subeq:init_C}, with $\delta=2h(0)$ since $\eta(x,y)=0$. This gives a molecular diffusion growth rate equal to $\gamma_D=3\kappa/h(0)^2$. In non-dimensional form at a unit Schmidt number, this is $\gamma_D^\star(\mathsf{R,S,D})=12\,[2h(0)k_0]^{-2}\mathsf{R}^{-4/3}$. A diffusive case is thus characterized by a small initial Reynolds $\mathsf{R}$, or by an effective diffusive thickness number $(2h(0)k_0)$ sufficiently large in comparison. This allows to determine the initial conditions at which diffusion becomes predominant, i.e. $\gamma^\star_D\geq\gamma_0^\star$, as those having an initial Reynolds number that verifies
\begin{equation}\label{eq:diffusion_vs_buoyancy_initial_R}
	\mathsf{R}^2\leq \psi\left(12\,[2h(0)k_0]^{-2}+2\right)^2-\psi.
\end{equation}

\begin{figure}[t!]
	\centering
	\includegraphics[width=\linewidth]{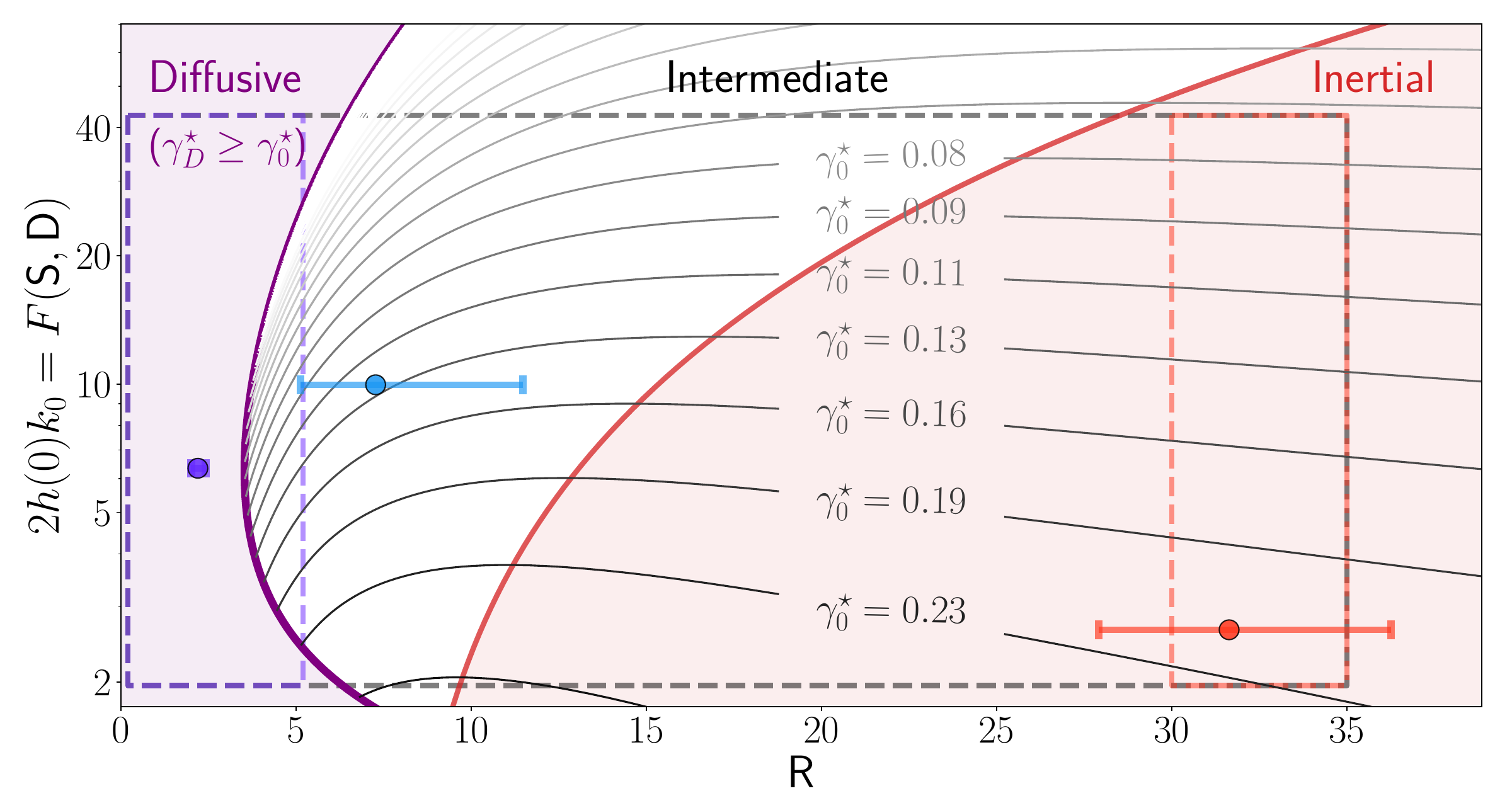}
	\caption{Positions of the three examples in the stability diagram, with the horizontal error bars indicating the extents of the spectral bands. The red and purple lines respectively correspond to Eqs. \eqref{eq:most_unstable_initial_R} $\&$ \eqref{eq:diffusion_vs_buoyancy_initial_R}. The dashed rectangles correspond to the domains in which global sensitivity analyses are performed in Section \ref{sec:late_time_dependence}.}
	\label{fig:examples_stability_diagram}
\end{figure}

In the end, Eqs. \eqref{eq:most_unstable_initial_R} $\&$ \eqref{eq:diffusion_vs_buoyancy_initial_R} allow to draw a stability diagram, shown in Figure \ref{fig:examples_stability_diagram}, which maps the typical behaviors as a function of the initial conditions. To be complete, we can even place the edges of the perturbation spectral band by replacing $\mathsf{R}$ with $\mathsf{R}_{\pm}=\mathsf{R}(1\pm \frac{1}{2}\mathsf{B})^{-3/2}$. Note that $\mathsf{R}_+$ corresponds to the smallest wavenumber mode of the spectral band, $(k_0-\frac{1}{2}\Delta k)$, while $\mathsf{R}_-$ corresponds to the largest, $(k_0+\frac{1}{2}\Delta k)$. 

\section{Late-time dependency on the initial conditions}\label{sec:late_time_dependence}

To explore the influence of the initial conditions on the late-time flow properties in a quantitative way, we carry out uncertainty and global sensitivity analyses \citep{saltelli_global_2007,saltelli_variance_2010,saltelli_why_2019} in three different domains, as indicated in Figures \ref{subfig:examples_init_nondim} $\&$ \ref{fig:examples_stability_diagram}. Two of them have the same volume and focus on the diffusive and inertial zones of the stability diagram, while the third is an extended domain spanning the whole range of initial conditions.

Performing these analyses in three different domains helps to focus on one type of behavior at a time, and to overcome the bounded nature of the analysis, as it depends on the size of the numerical domain. Indeed, an input factor, here one of the four parameters $i\in\mathsf{I}=(\mathsf{R,B,S,D})^T$ characterizing the initial conditions, can greatly contribute to the variance of the dynamics either because it is highly influential, or because it varies a lot. By comparing the results in the different domains, it is possible to identify overall trends and generalize the results with the help of theoretical considerations.

For each domain, tens of thousands of sets $\mathsf{I}$ of the four non-dimensional numbers are drawn randomly following uniform distributions for $\mathsf{R}$ and $\mathsf{B}$ and log-uniform distributions for $\mathsf{S}$ and $\mathsf{D}$. The dynamics associated to each set of initial conditions is computed with the surrogate model and the late-time properties are evaluated by taking the average over the time window $t^\star\in[160,170]$. This is late enough for the quantities to have converged to their asymptotic value and yet not too late to prevent the surrogate from extrapolating too much. We have verified that the conclusions remain unchanged by varying the time window.

\subsection{Uncertainty analysis}\label{subsec:uncertainty_analysis}

\begin{figure}[t!]
	\centering
	\includegraphics[width=\linewidth]{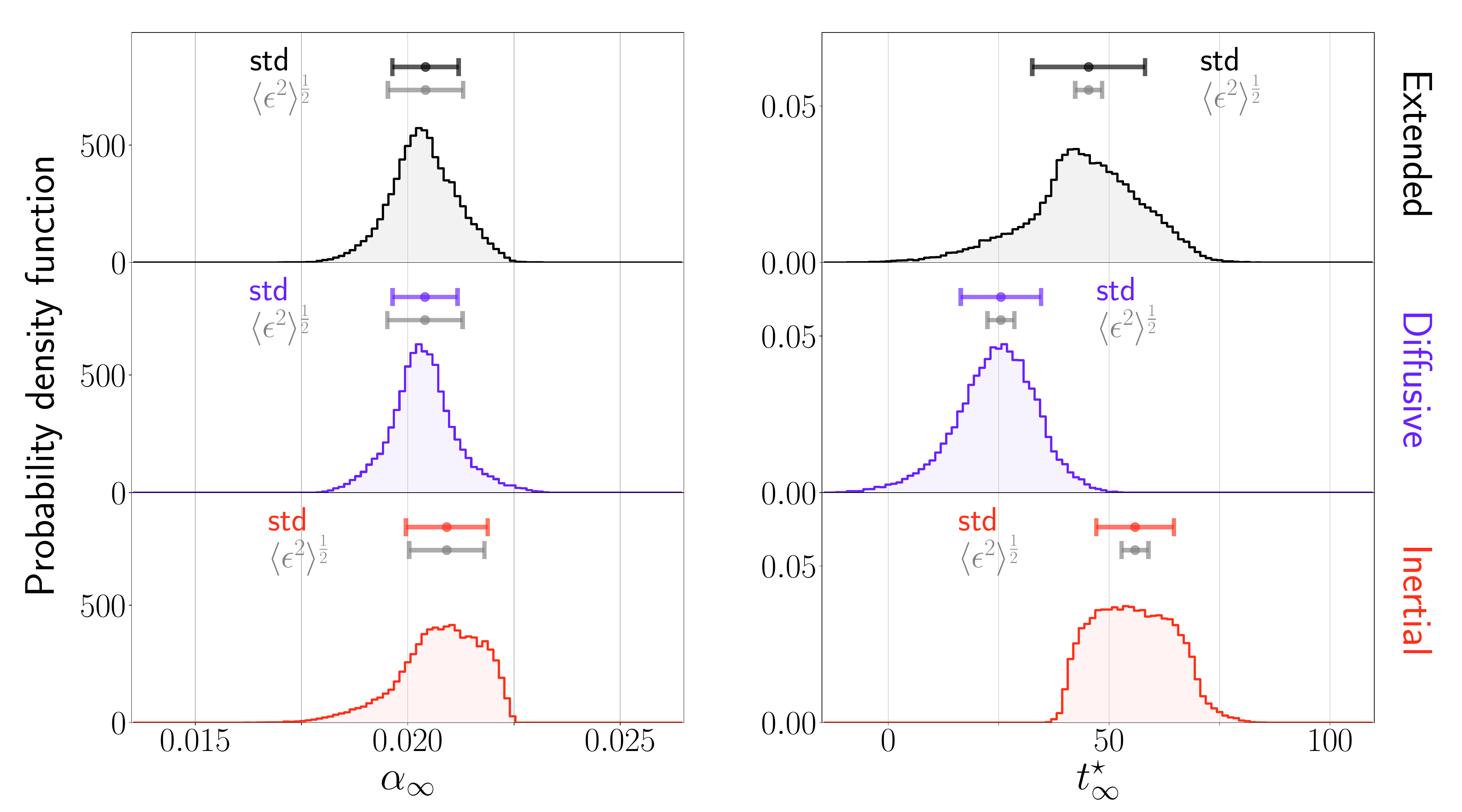}
	\caption{Probability density function (p.d.f.) estimates of the late-time properties describing the growth of the mixing layer. These are obtained by randomly drawing $4\times 10^4$ pairs of the four non-dimensional numbers describing the initial conditions in each domain, and computing the associated late-time dynamics with the surrogate model as the average over the time window $t^\star\in[160,170]$. The median value of the p.d.f. is indicated by the circle marker in the center of the horizontal bars. The upper bars represent plus or minus the standard deviation of the p.d.f. (std), whereas the lower grey bars represent plus or minus the square root of the surrogate model global error, $\langle\epsilon^2\rangle^{\frac{1}{2}}$ (see \ref{appendix:surrogate_approx_error}).}
	\label{fig:hists_late_time}
\end{figure}

Figure \ref{fig:hists_late_time} shows the probability density function (p.d.f.) estimates for the asymptotic growth rate $\alpha_\infty$ and virtual time origin $t^\star_\infty$. If the p.d.f. is very spread out, it means that the quantity highly depends on the initial conditions, and any prediction made without knowing them is highly uncertain. However, it is important to distinguish the part of the variance that is due to the surrogate's approximation errors from the part that comes from the physics. In Figure \ref{fig:hists_late_time}, the sprawl of the p.d.f. is represented by the upper horizontal bars, which correspond to twice the standard deviation (std), while the surrogate's global error is represented by the lower grey horizontal bars, which correspond to twice the square root of the measure defined in \ref{appendix:surrogate_approx_error}. If the latter is significantly smaller, then the quantity strongly depends on the initial conditions. In the opposite case, the p.d.f. mainly reflects the approximation errors of the surrogate.

The first striking result of this uncertainty analysis is that the self-similar growth rate varies very little in all domains. In particular, its variance is close to the error of the surrogate, and far from the wide range of values reported in the literature, that are roughly between $0.020$ and $0.070$ \citep{Dimonte_alpha_group_2004}. Our results therefore suggest that the asymptotic growth rate does not depend on the initial conditions, or too little to be detected with the surrogate. The median value is approximately $0.020$ in all three domains, which is consistent with most recent simulations \citep{cabot_reynolds_2006,morgan_parametric_2020,briard_growth_2022} and well-controlled experiments \citep{roberts_effects_2016}.

In contrast, the virtual time origin varies considerably, from a median value of $\sim25.5$ in the diffusive zone to $\sim55.9$ in the inertial zone. This dependency can clearly be attributed to a flow property, as the error of the surrogate is much smaller than the p.d.f.'s variance.

\subsection{Global sensitivity analysis} \label{subsec:global_SA}

To determine which parameters describing the initial conditions have the greatest influence on the late-time properties of the mixing layer, a variance-based global sensitivity analysis is carried out in each domain \citep{saltelli_global_2007}.

Analysis of this kind leads to the calculation of the so-called Sobol indices, which quantify the effects that has an input factor $i$ or group of input factors on the quantities $q$ of interest. Here, the input factors are the four non-dimensional parameters $i\in\mathsf{I}$, and we are interested in the first-order and total Sobol indices, $\mathsf{s}_i^{q}$ and $\mathsf{s}_{Ti}^{q}$. The first represents the main effect that input factor $i$ has on $q$, while the second represents its total effect, including both the effects it has alone and through interactions with other input factors. A detailed introduction is provided in \ref{appendix:Sobol_indices}.

\begin{figure}[t!]
	\centering
	\includegraphics[width=\linewidth]{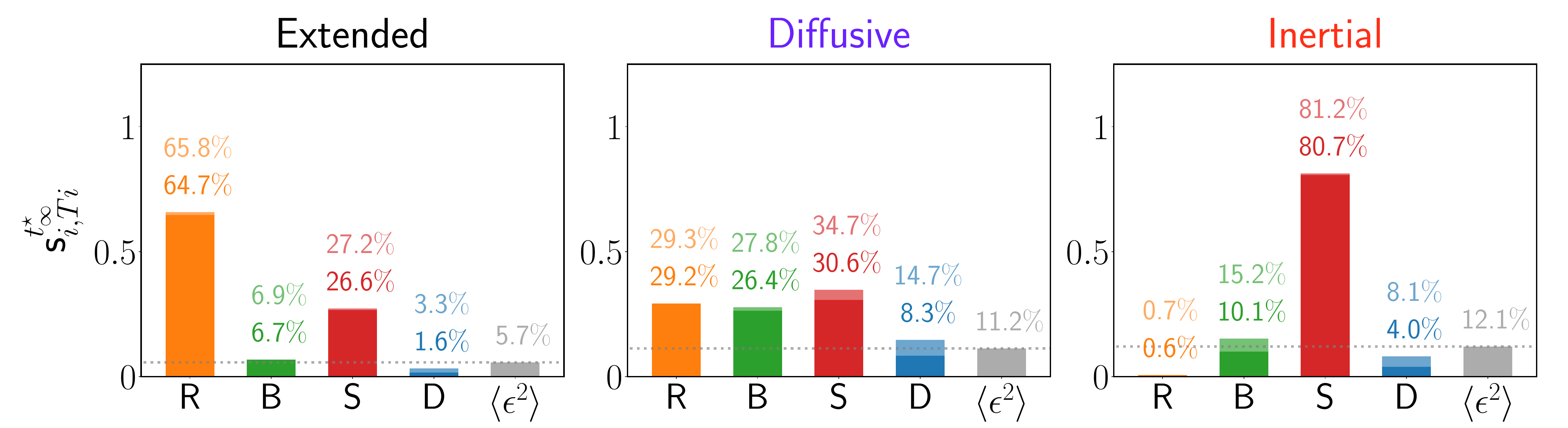}
	\caption{First-order (in full color) and total (in transparent color) Sobol indices, Eqs. \eqref{eq:Sobol_main_effect} $\&$ \eqref{eq:Sobol_total_effect}, of every input factor $i\in\mathsf{I}=(\mathsf{R,B,S,D})^T$ on the virtual time origin in the three domains. The indices are computed with the Monte Carlo estimates proposed by Jansen \citep{jansen_analysis_1999}, Eqs. \eqref{eqs:Monte_Carlo_estimates_Jansen1999}, where matrices $\mathsf{I}_A$ and $\mathsf{I}_B$ are each composed of $N=10^5$ samples. The global error $\langle\epsilon^2\rangle$ of the surrogate (see \ref{appendix:surrogate_approx_error}), divided by the total variance within the considered domain, is also shown for comparison.}
	\label{fig:Sobol_indices}
\end{figure}

As with the uncertainty analysis, only the parameters that affect the quantities significantly more than the surrogate's approximation error can be considered influential. In this respect, commenting on the Sobol indices for the asymptotic growth rate $\alpha_\infty$ is physically meaningless. 
Figure \ref{fig:Sobol_indices} therefore shows the main and total effects of the four initial conditions numbers only on the virtual time origin $t^\star_\infty$. This reveals its dependency on both the initial Reynolds number $\mathsf{R}$, the steepness number $\mathsf{S}$, the bandwidth number $\mathsf{B}$ and, to a lesser extent, the diffusive thickness number $\mathsf{D}$. As we have already seen with the uncertainty analysis, its value increases substantially between the diffusive and inertial zones, which indicates a strong initial Reynolds dependency that is confirmed by the sensitivity analysis in the extended domain.

In the diffusive domain, a perturbation having a large bandwidth number is far more likely to have RT unstable modes than one having a small bandwidth number, which may have only linearly stable modes. This difference may lead to two completely different behaviors, as discussed in Section \ref{sec:behaviors_unbounded_RT}, that can explain the influence of $\mathsf{B}$ in this domain. The sensitivity to the steepness number $\mathsf{S}$, which is particularly strong in the inertial zone, can be explained in two ways. First, it has a major influence on the initial size of the mixing layer $h_0$, as depicted by Eq. \eqref{eq:initial_mixing_zone_size_h0}, such that a large steepness number leads to a large initial size and consequently to a larger virtual origin. Second, a large steepness number accelerates the onset of nonlinear saturation and thereby the transition to turbulence, which may affect the virtual origin. This is explored in more detail in Section \ref{sec:model}.

\subsection{Focus on the inertial domain} \label{subsec:focus_inertial_domain}


In the inertial limit, viscous and diffusive effects are expected to play a marginal role in the dynamics. It is therefore reasonable to assume, as a first approximation, that the half size $h(t)$ of the mixing zone depends neither on the kinematic viscosity nor on the transport coefficient of molecular diffusion. It is thus a function $f$ of the time ($t$), the characteristics of the initial interface ($k_0$, $\Delta k$, $\eta_0$, $\delta$) and the buoyant acceleration ($\mathcal{A}g$). We can thus write
\begin{equation}\label{eq:scaling_inertial_trajectory}
	\begin{aligned}
		k_0 \,h = \mathsf{R}^{-2/3}\,h^\star &= f(\sqrt{\mathcal{A}gk_0}\,t, \Delta k/k_0, \eta_0 k_0, \delta k_0)\\
		&= f(\mathsf{R}^{-1/3}t^\star,\mathsf{B},\mathsf{S},\mathsf{D})
	\end{aligned}
\end{equation}
Consequently, any inertial trajectory, indexed $a$, can be expressed as a function of another, indexed $b$, as
\begin{equation}\label{eq:scaling_inertial_trajectory_fct_of_another}
	h_a^\star(t^\star,\mathsf{R}_a,\mathsf{B,S,D}) = \left(\mathsf{R}_a/\mathsf{R}_{b}\right)^{2/3} h_b^\star\left((\mathsf{R}_{b}/\mathsf{R}_a)^{1/3} t^\star,\mathsf{R}_{b},\mathsf{B,S,D}\right),
\end{equation}
where both trajectories share the same numbers $\mathsf{B}$, $\mathsf{S}$ and $\mathsf{D}$, but have different initial Reynolds. Trajectory $a$ has amplitude and time shifts with respect to trajectory $b$, which are modulated by power laws of their initial Reynolds number ratio. It is worth pointing out that this ratio is equal to the ratio of the classical growth rates of their mean wavenumber if the two trajectories have the same viscosity. Nonetheless, the scaling law also works if viscosities are different, which extends its use considerably. A qualitative check of Eq. \eqref{eq:scaling_inertial_trajectory_fct_of_another} is given in \ref{appendix:inertial_scaling_law}.

This scaling law has interesting implications. If we inject the self-similar solutions, Eq. \eqref{eq:self_similar_solution}, we find by identification that we must have
\begin{subequations}
	\begin{align}
		&\alpha_{\infty,a} = \alpha_{\infty,b} \\
		\text{and}\quad &t^\star_{\infty,a} = \left(\mathsf{R}_a/\mathsf{R}_{b}\right)^{1/3} t^\star_{\infty,b}.
	\end{align}
\end{subequations}
The asymptotic growth rate of both trajectories must be equal, which is consistent with the results of both the uncertainty and sensitivity analyses, as it does not depend on the initial Reynolds number. It could, however, depend on the three other numbers. The interesting point concerns the virtual time origin, which varies as a $1/3$ power-law function of the initial Reynolds, thus explaining the dependency found with the sensitivity analysis.

\begin{figure}[t!]
	\centering
	\includegraphics[width=\linewidth]{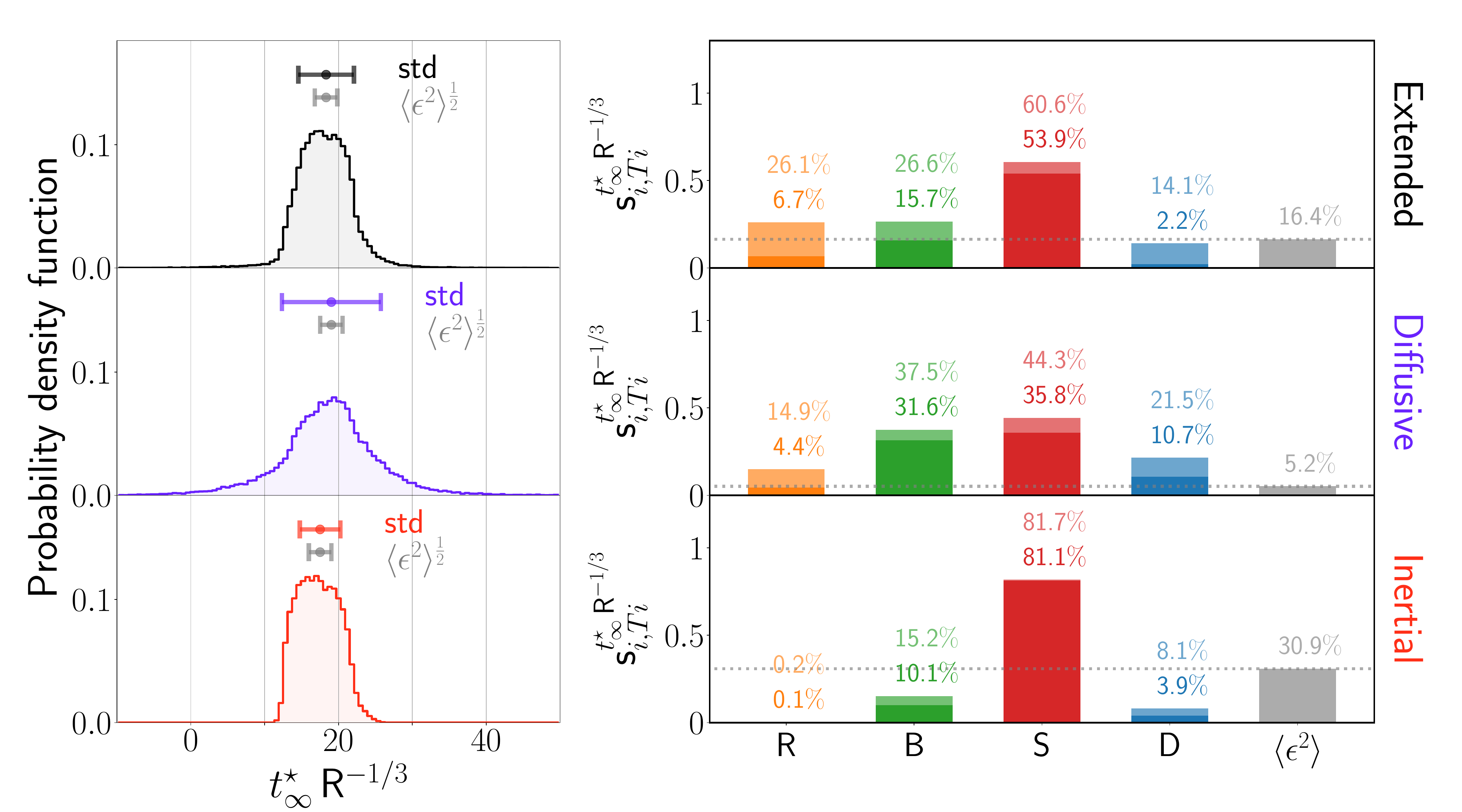}
	\caption{Uncertainty (left) and sensitivity (right) analyses for the renormalized virtual time origin. On the left are shown its probability density functions (p.d.f.) in the three domains, and on the right are shown the associated first-order (full color) and total (transparent color) Sobol indices. The p.d.f. are computed with $4\times10^4$ randomly drawn samples whereas the indices are computed with $N=10^5$ samples. The global error $\langle\epsilon^2\rangle$ of the surrogate is also shown for comparison (see \ref{appendix:surrogate_approx_error}).}
	\label{fig:UA_SA_tstarInfRpm13}
\end{figure}
To verify this in a quantitative way, we performed the uncertainty and sensitivity analyses again, but this time on the renormalized quantity $t^\star_\infty\,\mathsf{R}^{-1/3}$, which can also be written $t_\infty\,\gamma_0$, with $\gamma_0=\sqrt{\mathcal{A}gk_0}$ the classical RT growth rate of the perturbation's mean wavenumber. Figure \ref{fig:UA_SA_tstarInfRpm13} shows the results. It can be seen on the left that all p.d.f. collapse in the same range of values, with a median $\sim18$. The variance remains significantly larger than the surrogate error. Surprisingly, the inertial scaling law works reasonably well in the diffusive domain, although the assumptions are far from justified. The diffusive p.d.f. has nonetheless heavy tails and a variance around three times greater than that in the extended domain.

The sensitivity analysis, on the right, confirms that the main effect of the initial Reynolds is well captured by the $1/3$ power-law scaling, as the first-order Sobol index (in orange) is close to zero and inferior to the surrogate error in all domains. However, the total Sobol index reveals that the initial Reynolds is also involved in influential interactions with other factors in the diffusive domain, especially $\mathsf{B}$ and $\mathsf{D}$ that also have total indices larger than their first-order indices. These bandwidth and diffusive thickness numbers seem to be responsible for the large tails of the diffusive p.d.f., as only the steepness number is influential in the inertial domain. By removing the main effect of the initial Reynolds, we are able to refine the other dependencies. While $\mathsf{B}$ and $\mathsf{S}$ were already clearly identified as influential in Figure \ref{fig:Sobol_indices}, the influence of $\mathsf{D}$ is here confirmed. This may come from its impact on both the RT and molecular diffusion growth rates, following Eq. \eqref{eq:linear_growth_rate_nondim} and $\gamma_D^\star=12[2h(0)k_0]^{-2}\mathsf{R}^{-4/3}$, or on the initial size of the mixing zone, according to Eq. \eqref{eq:initial_mixing_zone_size_h0}.

\begin{figure}[t!]
	\centering
	\includegraphics[width=\linewidth]{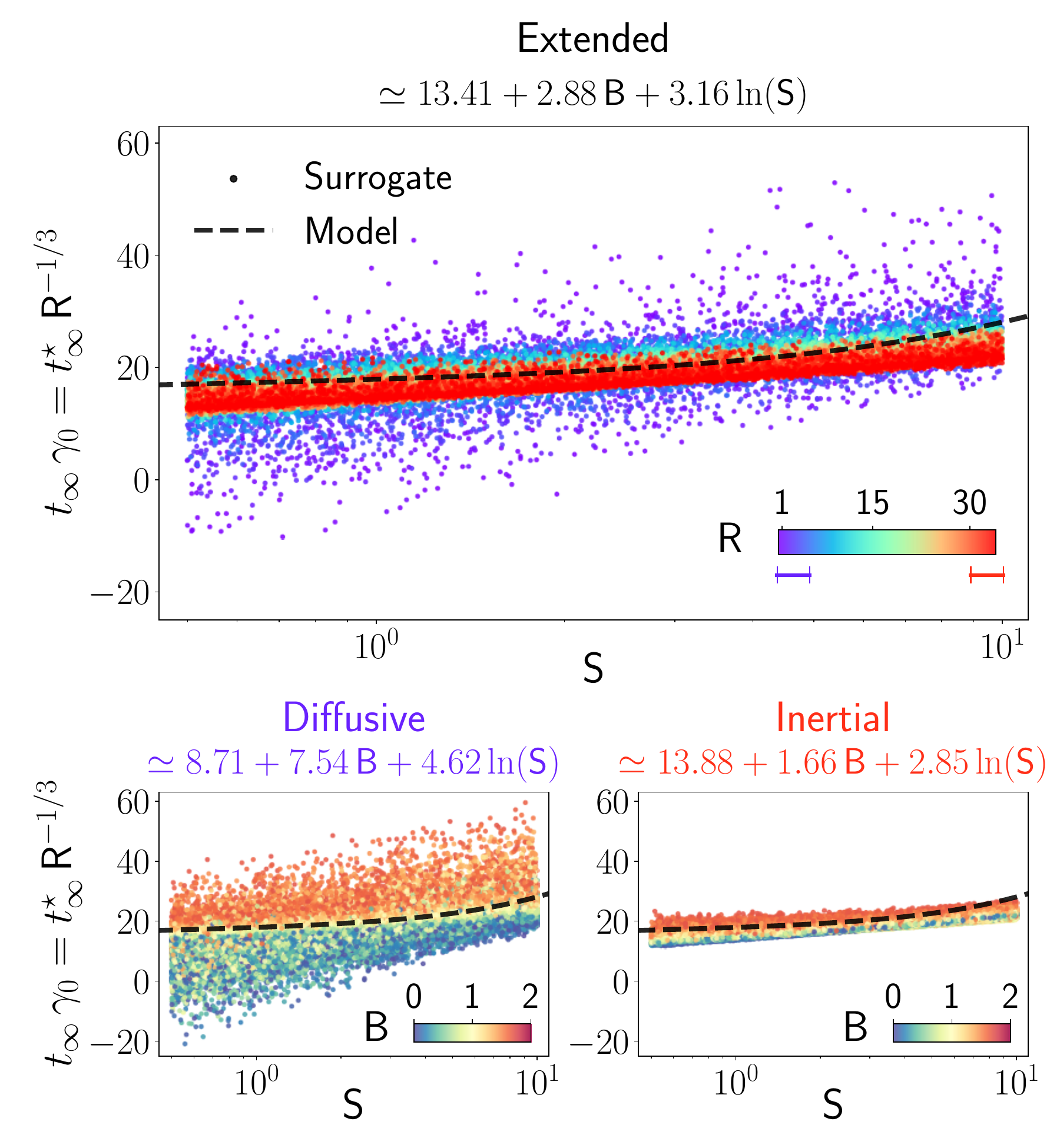}
	\caption{Logarithmic dependency of the renormalized virtual time origin on the steepness number $\mathsf{S}$. Each marker corresponds to the estimate computed with the surrogate model for a set of initial conditions $\mathsf{I}=(\mathsf{R,B,S,D})^T$, randomly drawn in the extended (top), diffusive (bottom left) and inertial (bottom right) domains. Linear regressions are given above the panels and the analytical model described in Section \ref{sec:model}, Eq. \eqref{eq:model_tinf_gamma0}, is shown in dashed black lines, using $h(0)=3\eta_0/2$, $\sigma=(\pi\sqrt{2(1+\mathcal{A})}\,)^{-1}$ and $\alpha_\infty=0.020$.}
	\label{fig:linear_fit}
\end{figure}

Summing the first-order indices whose associated variance is greater than or roughly equal to the estimated error of the surrogate model in the extended domain, thus only $\mathsf{B}$ and $\mathsf{S}$, we find that the main effects explain $69.6\%$ of the variance of $t_\infty\,\gamma_0=t^\star_\infty\,\mathsf{R}^{-1/3}$. In the inertial domain, this percentage reaches $81.1\%$, entirely due to the steepness number $\mathsf{S}$. By randomly drawing $4\times 10^4$ samples in the three domains and computing the associated renormalized virtual time origin with the surrogate, we find a roughly logarithmic dependency on the steepness number, as shown in Figure \ref{fig:linear_fit}. The dispersion around this trend is due to the bandwidth number, as indicated by the color bar, and is particularly large in the diffusive domain. Results of linear regressions performed on the random samples are given above the figure frames, and the analytical model derived in the next section is shown in comparison to the samples.

\section{A simple phenomenological model}\label{sec:model}

So far, we have quantified the sensitivity of the virtual time origin to initial conditions, but without giving any physical interpretation. 
This section aims to provide one by constructing a simple phenomenological model. The main objective is to establish a link between the late-time evolution of the mixing layer and the early stages of the instability, in order to generalize the numerical results of the previous section.

The model relies on the primary assumption that the flow is driven by a single dominant scale, although multiple modes are involved. Since the initial disturbance consists of a unique spectral band, this dominant scale is chosen as the mean wavenumber $k_0$. This choice could be different if the perturbation included more than one spectral band \citep{banerjee_3d_2009,rollin_generating_2013}, or if it had a more complex shape. If, in these cases, interactions between modes proved to be significant, more sophisticated models could also be used. For instance, Haan developed a second-order nonlinear mode-coupling model \citep{Haan_weakly_1991} capable of predicting triadic interactions and the generation of new unstable modes. He also proposed a criterion for the onset of nonlinear saturation in multi-mode perturbations \citep{Haan_onset_1989}. These were revisited and extended by Ofer \textit{et al.} \citep{Ofer_modal_1996} to explore the loss of memory of initial conditions, by Dimonte \citep{Dimonte_dependence_2004} to model their influence on the self-similar growth rate and by Rollin and Andrews \citep{rollin_generating_2013} to develop a method for initializing turbulence models. Thévenin's thesis \citep{thevenin_contrib_2024} also presents two examples in which Haan's models faithfully reproduce the appearance of spikes at high density contrast and provide a first approximation of the backscattering phenomenon.

\subsection{Model description}

The mixing layer first grows exponentially following the solution of linear stability theory for the dominant scale \citep{Chandrasekhar1961}, with its half size evolving as
\begin{equation}\label{eq:evolution_linear}
	h(t)=h(0)\cosh(\gamma_0\,t).
\end{equation}
The initial value $h(0)$ at $t=0$, given by Eq. \eqref{eq:initial_mixing_zone_size_h0}, is roughly $h(0)\simeq 3\eta_0/2$ in the inertial limit, such that $h(0)k_0\simeq 3\mathsf{S}/2$. 
In addition, the linear growth rate of the mean wavenumber is $\gamma_0=\sqrt{\mathcal{A}gk_0}$.

Nonlinear saturation is often said to occur when the perturbation's characteristic height becomes comparable to its characteristic wavelength \citep{lewis_instability_1950,sharp_overview_1984}. 
Saturation would therefore happen when $h(t_s)\sim 2\pi\sigma/k_0$, with $\sigma$ a free parameter threshold discussed in Section \ref{subsec:model_parameter_estimation}. However, since viscous and pressure drags are what causes saturation, here we rather define the criterion as when the velocity reaches $\dot{h}(t_s)=2\pi\sigma\gamma_0/k_0$.
By assuming a smooth transition between the linear and nonlinear regime of the RT instability \citep{birkhoff_taylor_1954,cherfils_simple_1996,Dimonte_dependence_2004}, we can define the nonlinear saturation time as
\begin{equation}\label{eq:nonlinear_saturation_time}
	t_s=\dfrac{1}{\gamma_0} \text{arcsinh}\left(\dfrac{2\pi\sigma}{h(0)k_0}\right).
\end{equation}
This time tends to zero as the effective steepness number $h(0)k_0$ increases, meaning that the onset of nonlinear saturation occurs earlier. It is worth pointing out that the inverse hyperbolic sinus is close to a logarithm, as $\text{arcsinh}(a)=\ln(a+\sqrt{1+a^2})$. In particular, it tends to $\ln(2a)$ when $a\gg1$, which is when the steepness number is small, i.e. $h(0)k_0\ll 2\pi\sigma$ in Eq. \eqref{eq:nonlinear_saturation_time}. 

\begin{figure}[t!]
	\centering
	\includegraphics[width=\linewidth]{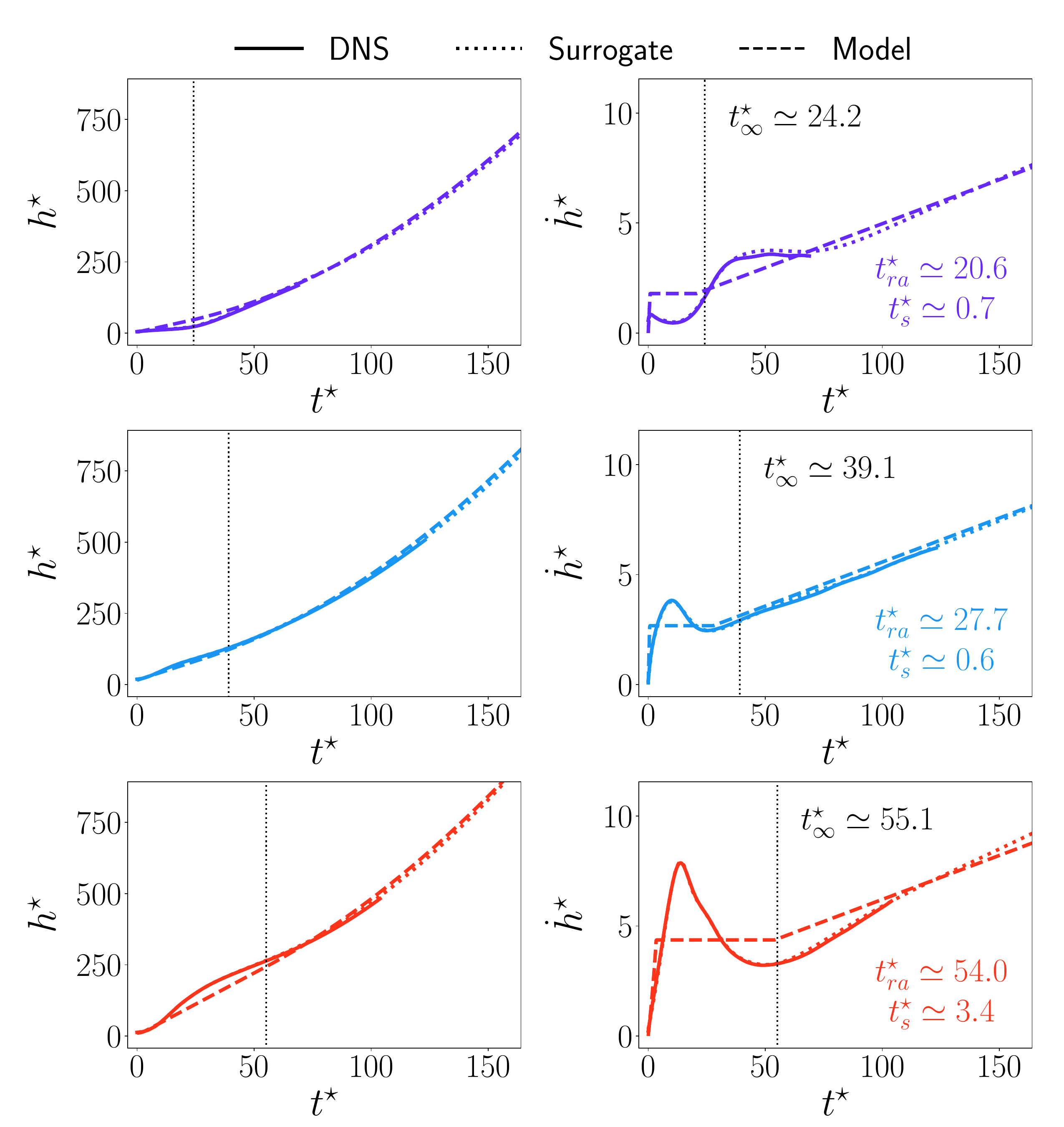}
	\caption{Comparison between the three DNS examples (solid lines) of Section \ref{sec:behaviors_unbounded_RT}, the surrogate predictions (dotted lines) and the model (dashed lines) described by Eqs. \eqref{eq:evolution_linear} to \eqref{eq:evolution_self_similar}, using $h(0)=3\eta_0/2$, $\sigma=(\pi\sqrt{2(1+\mathcal{A})}\,)^{-1}$ and $\alpha_\infty=0.020$.}
	\label{fig:model_vs_DNS}
\end{figure}

After nonlinear saturation, we assume that the mixing layer evolves at a constant velocity $V_0=\dot{h}(t_s)$ until re-acceleration at $t=t_{ra}$, similarly to what happens with single-mode perturbations. In addition, we assume that during this time period ($t_s\leq t\leq t_{ra}$), the dominant wavenumber remains $k_0$. Although not true, given that the trajectory may exhibit a local velocity peak with non-monotonic behavior and that the dominant wavenumber may decrease earlier, this correctly reflects the velocity evolution to a first approximation (see Figure \ref{fig:model_vs_DNS}). By integrating the saturation speed between $t_s$ and $t>t_s$, this gives an evolution equal to
\begin{equation}\label{eq:evolution_at_saturation_speed}
	h(t\geq t_s)=h(t_s)+V_0(t-t_s),
\end{equation}
with $h(t_s)=h(0)\sqrt{1+(2\pi\sigma)^2/(h(0)k_0)^2}$ the half size at saturation. Note that this approaches $h(t_s)\rightarrow 2\pi\sigma/k_0$ when $h(0)k_0\ll 2\pi\sigma$.

Finally, the mixing layer ends up re-accelerating during the transition to turbulence, and its half size follows the self-similar solution \citep{ristorcelli_rayleightaylor_2004,cook_mixing_2004}, Eq. \eqref{eq:self_similar_solution}, reproduced below.
\begin{equation}\label{eq:evolution_self_similar}
	h(t\geq t_{ra})=\alpha_\infty\mathcal{A}g(t+t_\infty)^2
\end{equation}

Figure \ref{fig:model_vs_DNS} shows how the model described by Eqs. \eqref{eq:evolution_linear} to \eqref{eq:evolution_self_similar} compares against the three DNS taken as examples in Section \ref{sec:behaviors_unbounded_RT}, with the parameters $\sigma$, $t_{ra}$ and $t_\infty$ as defined in Section \ref{subsec:model_parameter_estimation}. Although very simple, the model is able to represent the trajectories fairly well. In particular, the self-similar solution is well aligned with the surrogate predictions.

\subsection{Parameters estimation} \label{subsec:model_parameter_estimation}

The threshold $\sigma$, defining the onset of nonlinear saturation, can serve as a free parameter for tuning the model. However, as it determines the saturation speed, we can also use alternate theories. Based on potential flow theory, several works have indeed successfully derived weakly nonlinear models describing the evolution of RT bubbles by specifying their shape. 
These models, pioneered by Layzer \citep{Layzer_instability_1955} and followed by many others \citep{hecht_potential_1994,mikaelian_analytic_1998,Goncharov_analytical_2002,sohn_simple_2003}, lead to terminal velocities. The adjective \say{terminal} refers to a brief, temporary equilibrium between buoyancy and drag, which roughly occurs at saturation. Goncharov's model \citep{Goncharov_analytical_2002}, for instance, leads in three dimensions to a velocity equal to
\begin{equation}\label{eq:Goncharov_velocity}
	V_0=\sqrt{\dfrac{2\mathcal{A}g}{(1+\mathcal{A})k_0}},
\end{equation}
which is higher for low-wavenumber modes and parameter-free. If we use this as saturation velocity, the threshold must be $\sigma=(\pi\sqrt{2(1+\mathcal{A})})^{-1}$. In the small Atwood limit ($\mathcal{A}=0$), it is equal to $\sigma\simeq 0.225$. This is consistent with the literature \citep{lewis_instability_1950,sharp_overview_1984}, which gives values typically ranging between $0.10$ and $0.40$, and close to $1/(2\pi)\simeq 0.16$ that corresponds to $h(t_s)k_0=1$ in the small steepness number limit.

Using Goncharov's speed \eqref{eq:Goncharov_velocity} at saturation and $\alpha_\infty=0.020$, consistently with the results of Section \ref{sec:late_time_dependence}, it is possible to analytically determine the re-acceleration time $t_{ra}$ and virtual time origin $t_\infty$ by considering the continuity relations of the model for both $h(t)$ and its time derivative $\dot{h}(t)$. This ultimately leads to
\begin{subequations}
	\begin{align}
		&t_\infty=\dfrac{V_0}{4\alpha_\infty\mathcal{A}g}+\dfrac{h(t_{s})}{V_0}-t_{s} \label{eq:model_tinf} \\
		&t_{ra}=t_s+\dfrac{V_0}{4\alpha_\infty\mathcal{A}g}-\dfrac{h(t_{s})}{V_0} \label{eq:model_tra}
	\end{align}
\end{subequations}
The virtual time origin can be written in fully explicit and non-dimensional form as
\begin{equation}\label{eq:model_tinf_gamma0}
	\begin{aligned}
		t_\infty\,\gamma_0 &= \dfrac{\pi\sigma}{2\alpha_\infty}+\dfrac{1}{2\pi\sigma}h(t_{s})k_0-t_{s}\,\gamma_0 \\
		&= \dfrac{\pi\sigma}{2\alpha_\infty} +\dfrac{1}{2\pi\sigma}\sqrt{(h(0)k_0)^2+(2\pi\sigma)^2} -\text{arcsinh}\left(\dfrac{2\pi\sigma}{h(0)k_0}\right)
	\end{aligned}
\end{equation}

The model thus provides an analytical solution for the virtual time origin and an explanation of its sensitivity to the steepness number $\mathsf{S}$, as $h(0)k_0\simeq 3\mathsf{S}/2$ in the inertial limit. This dependency can be traced back to the onset of nonlinear saturation, and so the virtual time origin has a direct connection with the saturation time $t_s$. The model also gives an analytical solution for the time $t_{ra}$ through Eq. \eqref{eq:model_tra}. Re-acceleration thus occurs earlier if the initial perturbation saturates faster.


The solution of Eq. \eqref{eq:model_tinf} compares well with the surrogate predictions, as shown in Figure \ref{fig:linear_fit}. It works particularly well in the inertial and intermediate domains, which is logical given the assumptions used to build the model. Nevertheless, it also gives the general trend in the diffusive domain, although a large dispersion due to the bandwidth number is observed. A large $\mathsf{B}$ gives a greater chance to have at least one unstable mode in the initial spectral band. It also implies stronger nonlinear interactions, which can generate unstable modes of longer wavelength faster and thus predominate over diffusion sooner.

\section{Conclusion}

Late-time sensitivity to initial conditions in Rayleigh-Taylor (RT) turbulence is a long-standing question, with significant implications for modeling. In this paper, we address this issue using an approach that combines simulations, machine learning and theory.

As initial conditions are highly dimensional, we first simplify the problem by considering the influence of four non-dimensional numbers instead: an initial Reynolds ($\mathsf{R}$), a bandwidth ($\mathsf{B}$), a perturbation steepness ($\mathsf{S}$) and a diffusive thickness number ($\mathsf{D}$). These represent the statistical properties of initial conditions that describe random-phase multi-mode perturbations of a diffuse interface in density, with two fluids initially at rest and having a low density contrast. Using linear stability theory, we distinguish the RT dynamics between two typical behaviors -- diffusive and inertial, which overlap in an intermediate region of the non-dimensional stability diagram. As their name suggests, diffusive cases are initially dominated by molecular diffusion, while inertial cases are driven by buoyant forces from the start.

A physics-informed neural network, trained on a large database of direct numerical simulations (DNS) \citep{thevenin_al_JFM_2025,thevenin_database_2025}, forms the cornerstone of this study. Thanks to the physical constraints imposed during its training, this surrogate model can make realistic predictions in extrapolation, whether at very late times beyond the reach of the database's simulations, or for previously unseen initial conditions. This enables to estimate the late-time parameters that describe the self-similar regime, namely the asymptotic growth rate $\alpha_\infty$ and virtual time origin $t_\infty$. Besides, its predictions are very cheap, so many evaluations can be made quickly.

We take advantage of these features to identify the role of the initial conditions on the late-time dynamics. The results suggest that the asymptotic growth rate does not depend on the initial conditions, or too little to be detected within the accuracy of the surrogate. With a median value equal to $\alpha_\infty\simeq 0.020$, this is consistent with large simulations and well-controlled experiments. On the other hand, the virtual time origin is found to strongly depend on the initial conditions. A variance-based global sensitivity analysis reveals that the initial Reynolds and steepness numbers are the most influential parameters.

Finally, we propose a simple phenomenological model that represents the mixing layer dynamics in three distinct stages: exponential growth, constant-speed growth, and self-similar quadratic growth. Although this is a highly simplified representation of the flow in the multi-mode case, it compares fairly well with simulations. Besides, it provides an analytical relationship between the virtual time origin and the initial conditions, thus explaining its sensitivity. It reveals that the virtual time origin is closely connected to the onset of nonlinear saturation.

In addition to shedding light on a fundamental problem, these findings may open the way to identifying optimal initial conditions for delaying or speeding up turbulent mixing in practical applications. It also provides insights on how to initialize simulations to ensure that the flow has time to enter the self-similar regime before confining, which is handy if one wants to study fully developed turbulence. Beyond that, the results encourage further research on more complex configurations, such as high density contrast or spherical geometry, which are closer to applications and whose effects could affect both the virtual origin and growth rate. 

\vspace{0.75cm}
\noindent
\textbf{Acknowledgements.} The simulations were performed at the French TGCC center.

\vspace{0.5cm}
\noindent
\textbf{Declaration of interests.} The authors report no conflict of interest.

\vspace{0.5cm}
\noindent
\textbf{Data availability statement.} The data that support the findings of this study are openly available in the French Fluid Dynamics Database (F2D2) \citep{thevenin_database_2025}, at https://doi.org/10.57745/VTM1PN.

\vspace{0.5cm}
\noindent
\textbf{Authors ORCID.} S. Thévenin, https://orcid.org/0000-0002-2921-9706 ; B.-J. Gréa, https://orcid.org/0000-0001-7355-4790.

\newpage

\appendix

\section{Surrogate model approximation error} \label{appendix:surrogate_approx_error}

To be able to draw conclusions from the uncertainty and sensitivity analyses, it is essential to know how much confidence we can place in the estimation of the self-similar quantities with the surrogate model, and hence in the results derived from them. A classical measure of accuracy is the local squared error,
\begin{equation}\label{eq:squared_error}
	\epsilon^2(t^\star,\mathsf{I};q)=\left( q_{\,\text{DNS}}(t^\star,\mathsf{I}) - q(t^\star,\mathsf{I}) \right)^2,
\end{equation}
where a reference quantity $q_{\text{DNS}}(t^\star,\mathsf{I})$ from DNS is compared to the surrogate prediction $q(t^\star,\mathsf{I})$ at any time $t^\star$ along a trajectory that is parameterized by a set $\mathsf{I}$ of initial conditions. As we are investigating late-time properties and have no DNS data reaching the late time window under consideration, the surrogate error is computed in the transient on all test data (not seen during the learning phase) whose time exceeds $t^\star=50$. In addition, since the distribution of local squared errors is not Gaussian in general, the metric deemed most representative of the surrogate's performance on a single test trajectory is the median $\epsilon^2_m(\mathsf{I};q)$. Consequently, the global performance over all test examples is also taken as the median of $\epsilon^2_m(\mathsf{I};q)$, that we write $\langle \epsilon^2 \rangle_q$. 
Note that this global error is representative of the surrogate's performance in the whole area reachable with DNS at a resolution of $1024^2\times2048$ grid points, where the test data lie, and could be smaller or larger in localized regions of the parameter space.

\section{Sobol indices} \label{appendix:Sobol_indices}

A variance-based global sensitivity analysis consists in assessing how much the variance $V[q]$ of a quantity $q$ would decrease if an input factor, here $i\in\mathsf{I}$, could be fixed. This reduced variance is the conditional variance $V_\mathsf{J}[q|i]$, computed over all possible values of the remaining input factors $\mathsf{J}\subset\mathsf{I}$, with $i\neq j$, $\forall j\in\mathsf{J}$. If it is much smaller than the total variance $V[q]$, this means that the input factor $i$ strongly influences the value of the quantity $q$. To get a global measure of this reduction, we can take the expected mean $E_i[V_\mathsf{J}[q|i]]$ over all possible values of $i$ within the considered domain. Alternatively, the measure $V_i[E_\mathsf{J}[q|i]]$, that is equal to $V[q]-E_i[V_\mathsf{J}[q|i]]$ according to the law of total variance \citep{saltelli_global_2007}, is often preferred. The first-order Sobol index, defined as
\begin{equation}\label{eq:Sobol_main_effect}
	\mathsf{s}_i^q = \dfrac{V_i[E_\mathsf{J}[q|i]]}{V[q]}=1-\dfrac{E_i[V_\mathsf{J}[q|i]]}{V[q]},
\end{equation}
therefore quantifies the main effect that the input factor $i$ has on the quantity $q$. If this Sobol index is close to one, the input factor explains most of the variance on its own and is thus highly influential. If it is close to zero, it has no effect alone, but can still contribute to interactions. The sum of the input factors' main effects is always less than one, $\sum_{i\in\mathsf{I}}\mathsf{s}_i^q\leq 1$.

This approach is also applicable to groups of input factors, and the associated Sobol indices represent the effects of interactions between input factors. As these higher order indices are more expensive to estimate and because there are many of them, it is possible to measure the total effect of a factor $i$ instead, with the total Sobol index defined as
\begin{equation}\label{eq:Sobol_total_effect}
	\mathsf{s}_{Ti}^q = 1-\dfrac{V_\mathsf{J}[E_i[q|\mathsf{J}]]}{V[q]} = \dfrac{E_\mathsf{J}[V_i[q|\mathsf{J}]]}{V[q]},
\end{equation}
which includes its main effect and all the interaction effects it is involved in. Note that the effect of the interaction between two input factors $i$ and $j\neq i$, for instance, is counted twice, both in the total Sobol index of $i$ and in that of $j$. This is also true for higher order interactions. Consequently, the sum of the input factors' total effects is always superior or equal to one, $\sum_{i\in\mathsf{I}}\mathsf{s}_{Ti}^q\geq 1$. In addition, the effects of all interactions can be computed as $1-\sum_{i\in\mathsf{I}}\mathsf{s}_i^q$, and the effects of all interactions involving $i$ as $\mathsf{s}_{Ti}^q-\mathsf{s}_i^q$.

These Sobol indices can be computed efficiently with Monte Carlo methods by recognizing that the estimates of the variances in the numerators of Eqs. \eqref{eq:Sobol_main_effect} $\&$ \eqref{eq:Sobol_total_effect} are equal to the covariance of two carefully designed matrices, see for instance \citep{sobol_sensitivity_1993,homma_importance_1996,jansen_analysis_1999,saltelli_variance_2010}. In this study, we use the estimates proposed by Jansen \citep{jansen_analysis_1999},
\begin{subequations}\label{eqs:Monte_Carlo_estimates_Jansen1999}
	\begin{align}
		&V_i[E_\mathsf{J}[q|i]]\simeq V[q]-\dfrac{1}{2N} \sum_{n=1}^{N} \left( q(\mathsf{I}_B^{(n)}) - q(\mathsf{I}_{A_B}^{(n)}) \right)^2 \label{subeq:Monte_Carlo_estimate_main_effect}\\
		\text{and}\quad &V_\mathsf{J}[E_i[q|\mathsf{J}]]\simeq \dfrac{1}{2N} \sum_{n=1}^{N} \left( q(\mathsf{I}_A^{(n)}) - q(\mathsf{I}_{A_B}^{(n)}) \right)^2, \label{subeq:Monte_Carlo_estimate_total_effect}
	\end{align}
\end{subequations}
where $\mathsf{I}_A$ and $\mathsf{I}_B$ are two matrices containing $N$ identically distributed and independent samples of $\mathsf{I}$, with each line corresponding to a sample and each column to an input factor. The matrix $\mathsf{I}_{A_B}$ is a copy of $\mathsf{I}_A$, except for the column corresponding to $i$, which has been replaced by the same column of matrix $\mathsf{I}_B$.

\section{Inertial scaling law}\label{appendix:inertial_scaling_law}

As a qualitative check of the scaling law in Eq. \eqref{eq:scaling_inertial_trajectory_fct_of_another}, we compare it with highly inertial DNS.

To reach even higher initial Reynolds numbers than those available in the database \citep{thevenin_database_2025}, without confining the large scales, three simulations were performed with a resolution of $2048^2\times 4096$ grid points. These DNS share the same values for most initialization parameters, with $k_0=30$, $\Delta k=6$, $\eta_0=\delta=0.0167$, $\nu=3\times 10^{-4}$ and $\mathcal{A}=0.05$, giving $\mathsf{B}=0.2$ and $\mathsf{S}=\mathsf{D}=0.5$. Only the accelerations $g$ differ, being respectively equal to $59.53$, $98.41$ and $147.0$, giving initial Reynolds numbers $\mathsf{R}$ equal to $35$, $45$ and $55$. Having small values for $\mathsf{S}$ and $\mathsf{D}$ guarantees to obtain inertial behaviors, and according to Eq. \eqref{eq:initial_mixing_zone_size_h0}, it gives a small initial size $h_0$. Consequently, it also delays lateral and vertical confinements of the large scales, as these appear when the mixing zone size $2h(t)$ reaches about the horizontal box size and a third of the vertical box size, respectively.

\begin{figure}[t!]
	\centering
	\includegraphics[width=\linewidth]{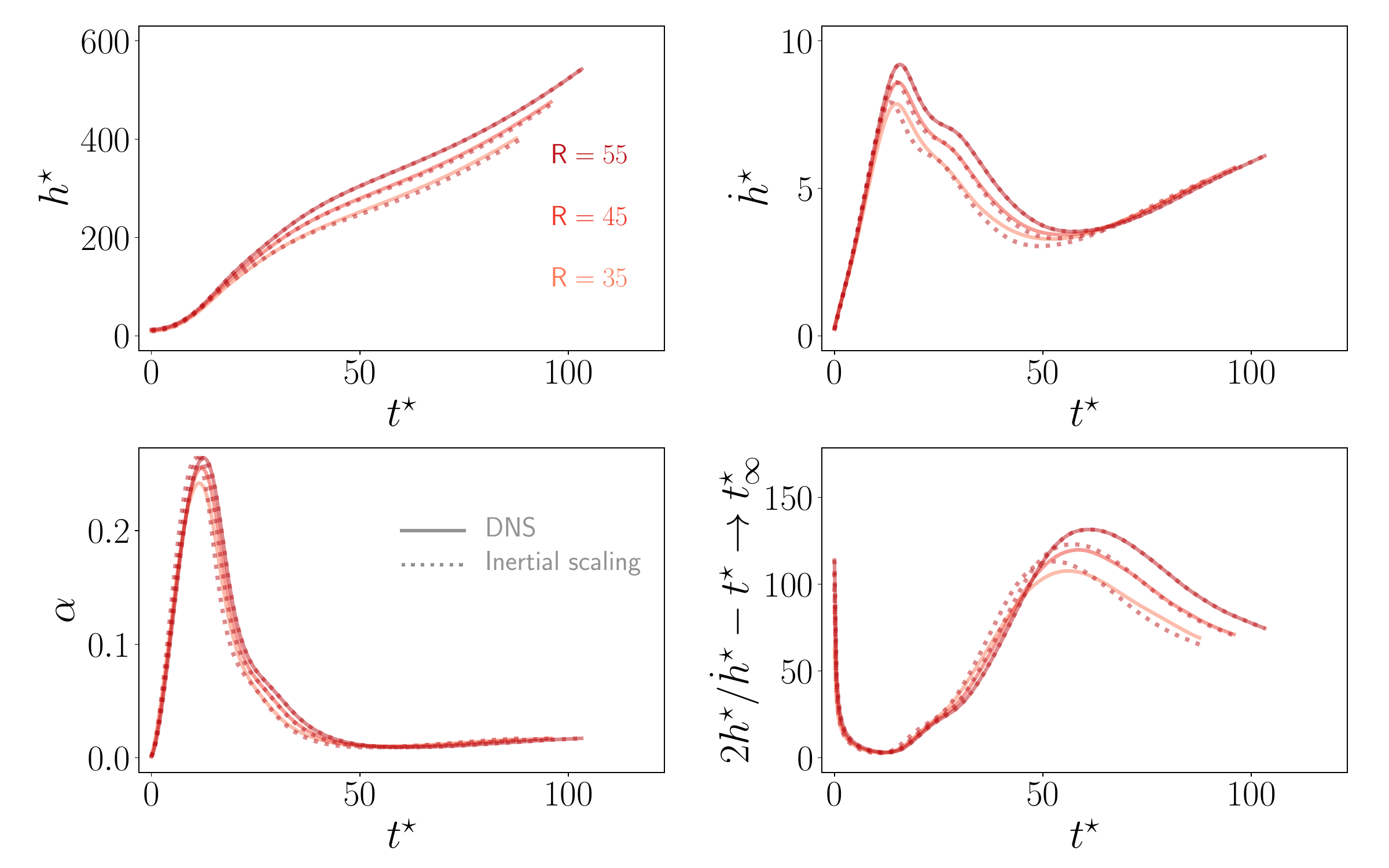}
	\caption{Three examples at high initial Reynolds number to illustrate the inertial scaling law in Eq. \eqref{eq:scaling_inertial_trajectory_fct_of_another}. These have the same numbers $\mathsf{B}$, $\mathsf{S}$ and $\mathsf{D}$, but initial Reynolds numbers $\mathsf{R}$ ranging from $35$ to $55$. Solid lines correspond to the DNS whereas dotted lines correspond to the inertial scaling, taking the DNS at $\mathsf{R}=55$ as reference.}
	\label{fig:inertial_invariance}
\end{figure}

Figure \ref{fig:inertial_invariance} shows the three simulations against the inertial scaling, where the DNS at $\mathsf{R}=55$ is taken as the reference trajectory, indexed $b$ in Eq. \eqref{eq:scaling_inertial_trajectory_fct_of_another}. The scaling is able to reproduce the time trajectories very well, although a small deviation is observable around $t^\star\sim 15$ at the peak of velocity, which corresponds to the development of secondary shear instabilities.

\newpage






\bibliographystyle{plain} 
\bibliography{biblio} 

\end{document}